\documentstyle[twoside,color,graphicx,fleqn,espcrc2]{article}
\def\eq#1{{eq.~(\ref{#1})}}
\def\eqs#1#2{{eqs.~(\ref{#1})--(\ref{#2})}}
\def\vev#1{\left\langle #1\right\rangle}
\def\abs#1{\left| #1\right|}
\def\mod#1{\abs{#1}}
\def\Im{\mathop{\mbox{Im}}}
\def\Re{\mathop{\mbox{Re}}}

\def\etal{{\it et al.}}

\def\qq{$\vev{\bar q q}$}
\def\GG{$\vev{\alpha_s GG/ \pi}$}
\def\eps{$\varepsilon$}
\def\eprime{$\varepsilon'$}
\def\ratio{$\varepsilon'/\varepsilon$}
\def\CP{$CP$}
\newcommand{\be}{\begin{equation}}
\newcommand{\ee}{\end{equation}}
\newcommand{\bea}{\begin{eqnarray}}
\newcommand{\eea}{\end{eqnarray}}
\newcommand{\nn}{\nonumber}

\begin{document}
%\draft
%%%%%%%%%%%%%%%%%%%%%%%%%%%%%%%%%%%%%%%%%%%%%%%%%%%%%%%%%%%  FRONT PAGE
\title{\ratio\ in the standard model}

\author{M.~Fabbrichesi\address{INFN, Sezione di Trieste and Scuola 
       Internazionale di Studi Superiori Avanzati,\\
       via Beirut 4, I-34014 Trieste, Italy.}%
       }%

\begin{abstract}\noindent
I discuss the estimate of the \CP-violating ratio \ratio\ by stressing
the role played by the chiral quark model in predicting the experiment
and in showing that the same dynamical mechanism at work in the
$\Delta I=1/2$ rule also explains the larger value obtained for
\ratio\ in this model with respect to other estimates.
\end{abstract}

%%%%%%%%%%%%%%%%%%%%%%%%%%%%%%%%%%%%%%%%%%%%%%%%%%%%%%%%%%%%%%%%%%%
\maketitle
%%%%%%%%%%%%%%%%%%%%%%%%%%%%%%%%%%%%%%%%%%%%%%%%%%%%%%%%%%%%%%%%%%%

\section{Introduction}

The \CP-violating  ratio \ratio\ (for a review see, e.g., 
\cite{WW,burasrev,review}) is computed as
\be
\Re\; \varepsilon '/\varepsilon = 
e^{i \phi}\ 
\frac{G_F \omega}{2\mod{\epsilon}\Re{A_0}} \:
\mbox{Im}\, \lambda_t \: \:
 \left[ \Pi_0 - \frac{1}{\omega} \: \Pi_2 \right] \; ,
\label{epsprime2}
 \ee
where, $\lambda_t = V_{td}V^*_{ts}$ is a combination of 
Cabibbo--Kobayashi-Maskawa (CKM) matrix elements,
\bea
 \Pi_0 & = &  
\frac{1}{\cos\delta_0} \sum_i y_i \, \Re \, \langle  Q_i  \rangle _0 
\ (1 - \Omega_{\eta +\eta'}) \; ,
\label{Pi_0}\\
 \Pi_2 & = & 
\frac{1}{\cos\delta_2} \sum_i y_i \, \Re \, \langle Q_i \rangle_2 \; ,
\label{Pi_2}
\eea
and $\langle  Q_i  \rangle _I =  \langle 2 \pi , I|  Q_i| K
\rangle$.
The  $\Delta S = 1$
effective lagrangian ${\cal L}_W$ is given by
 \be
 {\cal L}_{W}= - \sum_i  C_i(\mu) \; Q_i (\mu) \; ,
 \label{Lquark}
\ee
where
\be
C_i(\mu) =  \frac{G_F}{\sqrt{2}} V_{ud}\,V^*_{us} 
 \Bigl[z_i(\mu) + \tau\ y_i(\mu) \Bigr] \; .
\label{Lqcoef}
\ee
In (\ref{Lqcoef}), $G_F$ is the Fermi coupling, the functions $z_i(\mu)$ and
 $y_i(\mu)$ are the
 Wilson coefficients and $\tau = - V_{td}
V_{ts}^{*}/V_{ud} V_{us}^{*}$. 
According to the standard parameterization of the CKM matrix, in order to
determine \ratio\ , we only
need to consider the $y_i(\mu)$ components, 
which control the $CP$-violating part of the lagrangian.
The coefficients   $y_i(\mu)$, and $z_i(\mu)$ contain all the dependence
of short-distance
 physics, and depend on the $t,W,b,c$ masses, the intrinsic QCD scale
$\Lambda_{\rm QCD}$, the $\gamma_5$-scheme used in the regularization
and the renormalization scale $\mu$. They are known 
to the next-to-leading order (NLO) order in $\alpha_s$ and 
$\alpha_e$~\cite{NLO}.

The $Q_i$ in eq.~(\ref{Lquark})
 are the effective four-quark operators obtained 
in the standard model by integrating out 
the vector bosons and the heavy quarks $t,\,b$ and $c$. A convenient
and by now standard
basis includes the following ten  operators:
 \be
\begin{array}{rcl}
Q_{1} & = & \left( \overline{s}_{\alpha} u_{\beta}  \right)_{\rm V-A}
            \left( \overline{u}_{\beta}  d_{\alpha} \right)_{\rm V-A}
\, , \\[1ex]
Q_{2} & = & \left( \overline{s} u \right)_{\rm V-A}
            \left( \overline{u} d \right)_{\rm V-A}
\, , \\[1ex]
Q_{3,5} & = & \left( \overline{s} d \right)_{\rm V-A}
   \sum_{q} \left( \overline{q} q \right)_{\rm V\mp A}
\, , \\[1ex]
Q_{4,6} & = & \left( \overline{s}_{\alpha} d_{\beta}  \right)_{\rm V-A}
   \sum_{q} ( \overline{q}_{\beta}  q_{\alpha} )_{\rm V\mp A}
\, , \\[1ex]
Q_{7,9} & = & \frac{3}{2} \left( \overline{s} d \right)_{\rm V-A}
         \sum_{q} \hat{e}_q \left( \overline{q} q \right)_{\rm V\pm A}
\, , \\[1ex]
Q_{8,10} & = & \frac{3}{2} \left( \overline{s}_{\alpha} 
                                                 d_{\beta} \right)_{\rm V-A}
     \sum_{q} \hat{e}_q ( \overline{q}_{\beta}  q_{\alpha})_{\rm V\pm A}
\, , 
\end{array}  
\label{Q1-10} 
\ee
where $\alpha$, $\beta$ denote color indices ($\alpha,\beta
=1,\ldots,N_c$) and $\hat{e}_q$  are the quark charges 
($\hat{e}_u = 2/3$,  $\hat{e}_d=\hat{e}_s =-1/3$). Color
indices for the color singlet operators are omitted. 
The labels \mbox{$(V\pm A)$} refer to the Dirac structure
\mbox{$\gamma_{\mu} (1 \pm \gamma_5)$}.

Notice the explicit presence of the final-state-interaction (FSI)
phases $\delta_I$ in eqs.~(\ref{Pi_0}) and (\ref{Pi_2}). 
Their presence is a consequence of writing the absolute values 
of the amplitudes in term of their dispersive parts.

Finally, $\Omega_{\eta+\eta'}$ is the isospin breaking (for $m_u\neq
m_d$) contribution of the mixing of $\pi$ with $\eta$ and $\eta'$.

Experimentally the ratio \ratio\ is extracted, by collecting $K_L$ and
$K_S$ decays into pairs of $\pi^0$ and $\pi^\pm$, from the relation 
\be
\left|\frac{\eta_{+-}}{\eta_{00}}\right|^2 \simeq 
1 + 6 \; \Re\: \frac{\varepsilon '}{\varepsilon} \, ,
\ee
where
\be
\eta_{00} \equiv 
\frac{\langle \pi^0 \pi^0 | {\cal L}_W | K_L \rangle}
{\langle \pi^0 \pi^0 | {\cal L}_W | K_S \rangle}
\label{eta00}
\ee
and
\be
\eta_{+-} \equiv \frac{\langle \pi^+ \pi^- | {\cal L}_W | K_L \rangle}
{\langle \pi^+ \pi^- | {\cal L}_W | K_S \rangle} 
\label{eta+-} \, .
\ee

The announcement last year of the preliminary
result from the KTeV collaboration (FNAL)~\cite{KTeV}
based on data collected in 1996-97,
and the present result from the NA48 collaboration (CERN), 
based on data collected in 1997 and 1998~\cite{NA48},
settle the long-standing issue of the presence of direct \CP\ violation
in kaon decays. However, a clearcut determination 
the actual value of \eprime\ at the precision of a few
parts in $10^4$ must wait for further statistics
and scrutiny of the experimental systematics. 
By computing the average among the two 1992
experiments (NA31~\cite{NA31} and E731~\cite{E731}) 
and the KTeV and NA48 data we obtain
\be
\Re\: \varepsilon/\varepsilon' = (1.9 \pm 0.46) \times 10^{-3}\, , 
\label{average}
\ee
where the error has been inflated according to the Particle Data Group
procedure ($\sigma \to \sigma \times \sqrt{\chi^2/3}$), to be used when 
averaging over experimental data---in our case four sets---with 
substantially different central values.

The value in \eq{average} can be considered 
the current experimental result. Such a result
will be tested within the next year by the full data analysis
from KTeV and NA48 and (hopefully)
 the first data from KLOE at DA$\Phi$NE (Frascati); 
at that time, the experimental uncertainty will be
reduced to a few parts in $10^{4}$.

\section{Hadronic matrix elements}

If there is no sizable cancellation between
the relevant effective operators, the order of magnitude
of  $\varepsilon'/\varepsilon$ is bound to be of the order of
$10^{-3}$. A simple argument for this is presented
in~\cite{qcd99}. The problem is that any cancellation, or the lack thereof,
among the operators heavily 
depends on the size of the hadronic matrix elements and
there is no available estimate of them that is free of hard-to-control
 assumptions.
To estimate the hadronic matrix elements in a systematic
manner without having first to solve QCD (not even by lattice
simulation) we needed a 
model that is simple enough to understand its dynamics and, 
at the same time, not too simple so as to
still include the relevant physics. We chose the
chiral quark model ($\chi$QM)~\cite{trieste97} in which all coefficients of the
relevant chiral lagrangian are parameterized in terms of just three 
parameters: the quark and gluon condensates, and the quark constituent
mass. The model makes possible a complete estimate of all matrix elements, it
includes non-factorizable effects, chiral corrections and final-state
interaction, all of which we thought to be relevant.

 The $\chi$QM~\cite{QM} 
is an effective quark model of QCD which
can be derived in the framework of
the extended Nambu-Jona-Lasinio  model of chiral symmetry
breaking (for a review, see, e.g., \cite{ENJL}).
In the $\chi$QM an effective interaction between the $u,d,s$ quarks and the
meson octet is introduced via the term
\be
 {\cal{L}}_{\chi \mbox{\scriptsize QM}} = - M \left( \overline{q}_R \, \Sigma
\; q_L +
\overline{q}_L \, \Sigma^{\dagger} q_R \right) \ ,
\label{M-lag}
\ee 
which is added to an effective low-energy QCD 
lagrangian whose dynamical degrees
of freedom are the $u,d,s$ quarks propagating in a soft gluon background.
The matrix  $\Sigma$ in (\ref{M-lag}) is the same as that used in
chiral
perturbation theory and it 
contains the pseudo-scalar meson multiplet. The quantity $M$ is
interpreted as the constituent quark mass in mesons (current quark masses
are also included in the effective lagrangian).

The hadronic matrix elements are matched with the NLO Wilson coefficients
at the scale $\Lambda_\chi \simeq 0.8$ ($\simeq m_\rho$) as the
best compromise between the range of validity of chiral perturbation and
that of strong coupling expansion. 

\section{The fit of the $\Delta I =1/2$ rule}

In order to assign the values of the model-dependent parameters $M$,
\qq\ and \GG\ , we consider the $CP$-conserving amplitudes in
the $\Delta I = 1/2$ selection rule of $K\to\pi\pi$ decays. 
In practice, we compute the amplitudes
\bea
 A_0 & = &  \frac{G_F}{\sqrt{2}} V_{ud}\,V^*_{us} 
\frac{1}{\cos\delta_0} \sum_i z_i(\mu) \, \Re \langle Q_i(\mu) \rangle _0 \; ,
\label{A_0} \nn \\
 A_2 & = &  \frac{G_F}{\sqrt{2}} V_{ud}\,V^*_{us}
\frac{1}{\cos\delta_2} \sum_i  z_i(\mu) \, \Re \langle Q_i(\mu)
\rangle_2 \nn \\
& & + \omega\ A_0\ \Omega_{\eta +\eta'} \; ,
\label{A_2}
\eea
within the $\chi$QM approach and vary the parameters in order
to reproduce their experimental values.

This procedure combines a model for low-energy QCD---which allow us to
compute all hadronic matrix elements in terms of a few basic
parameters---with the phenomenological
determination of such parameters. In this way,
some shortcomings of such a naive model (in particular,
the matching between long- and short-distance components) are absorbed in the
phenomenological fit. As a check, we eventually verify
the stability of the computed observables against renormalization scale 
and scheme dependence.

The fit of the $CP$-conserving involves the determination of the 
FSI phases.  
The absorptive components of the hadronic matrix elements appear when
chiral loops are included.
In our approach the direct determination of the 
rescattering phases gives at $O(p^4)$ 
$\delta_0 \simeq 20^0$ and $\delta_2 \simeq -12^0$.
Although these results show features which are in qualitative
agreement with the phases extracted from pion-nucleon 
scattering the 
deviation from the experimental data is sizeable, especially in the
$I=0$ component. On the other hand, at $O(p^4)$ the absorptive parts of the
amplitudes are determined only at $O(p^2)$ and disagreement with the
measured phases should be expected.
As a matter of fact, the authors of ref.~\cite{dubna} find that
at $O(p^6)$ the absorptive part of the hadronic
matrix elements are substantially modified to give
values of the rescattering phases quite close to those 
extracted from pion-nucleon scattering.
At the same time the $O(p^6)$ corrections to the dispersive part
of the hadronic matrix elements are very small.  
This result corroborates our ansatz~\cite{trieste97} of trusting the
real parts of the $O(p^4)$ matrix elements while inputting
the experimental values of the rescattering phases
in all parts of our analysis, which amounts to taking
$\cos\delta_0 \approx 0.8$ and $\cos\delta_2 \approx 1$. 

%%%%%%%%%%%%%%%%%%%%%%% FIG.2 %%%%%%%%%%%%%%%%%
\begin{figure}
\includegraphics[scale=0.5]{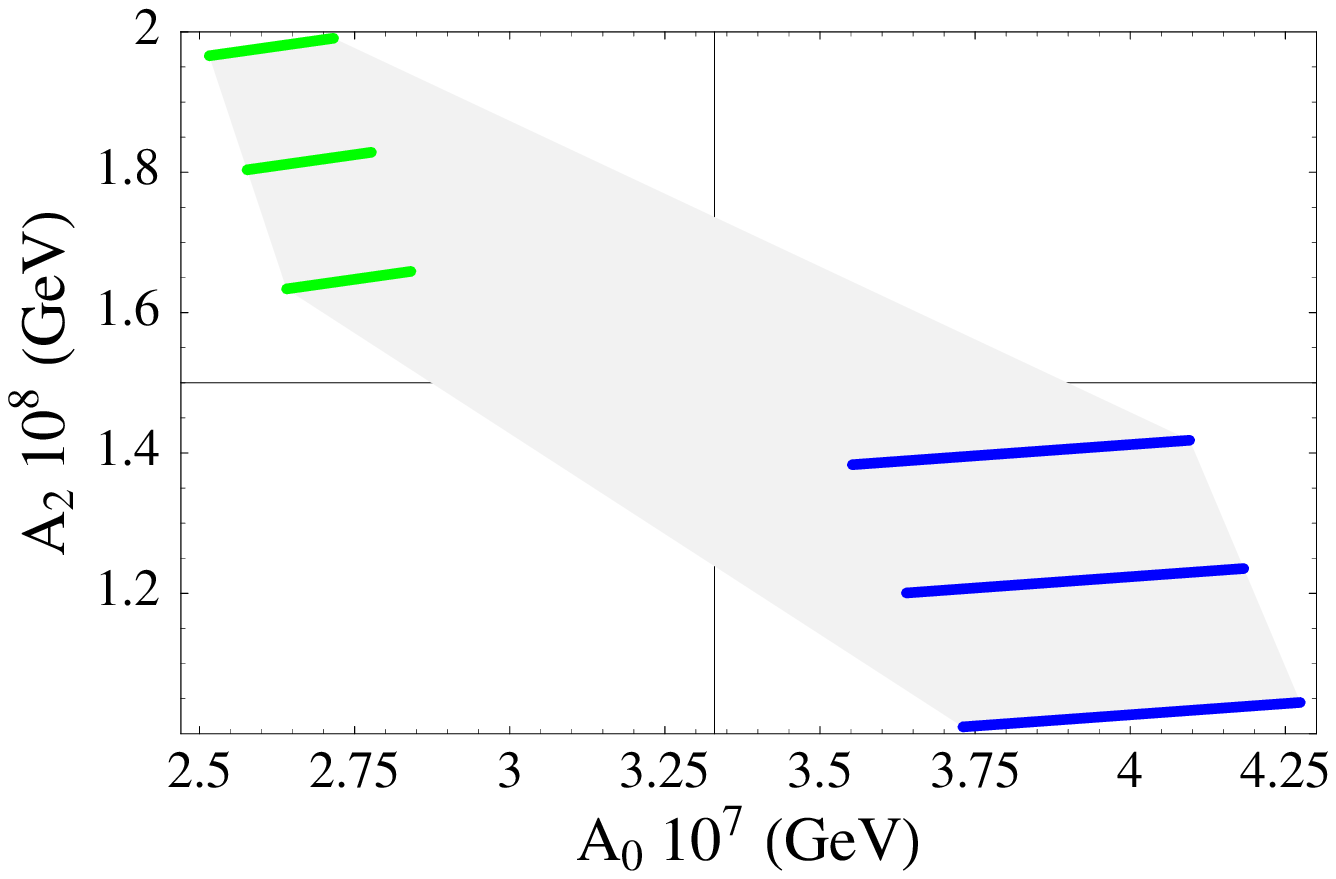}
\includegraphics[scale=0.5]{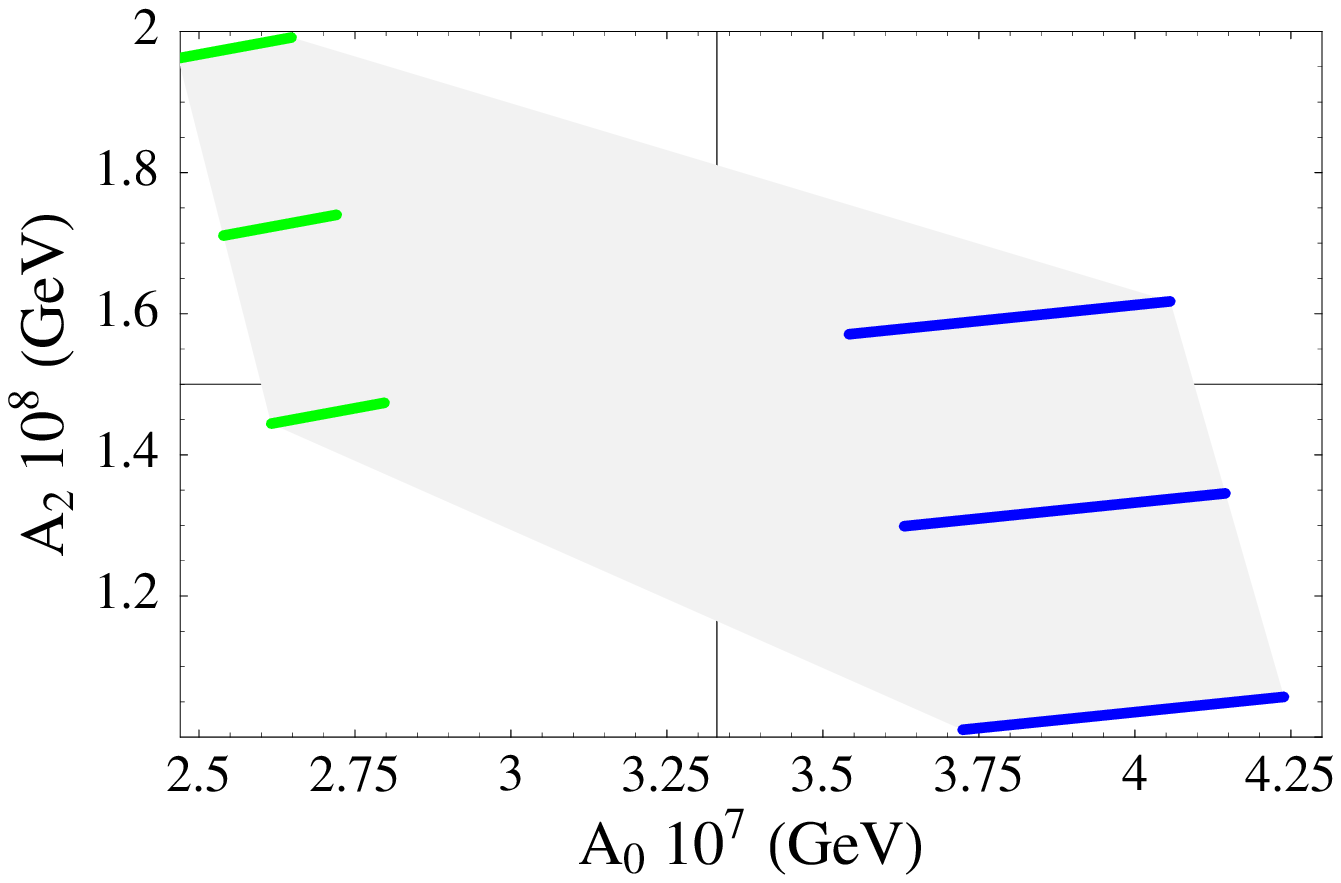}
\caption{Dependence of $A_0$ and $A_2$ on
$\vev{\bar{q}q}$, $\vev{GG}$, $\Lambda^{(4)}_{\rm QCD}$ 
and $M$ at $\mu = 0.8$ GeV.
The gray and black sets of lines correspond to the extreme values of 
$\Lambda_{\rm QCD}$ and $M$. 
The length of the lines represents the effect
of varying $\vev{\bar{q}q}$, while keeping all other parameters fixed.
The vertical spread corresponds to varying $\vev{\alpha_s GG/\pi}$, 
with the central line corresponding to the central value of $\vev{GG}$. 
The gray area denotes the region spanned by varying all the
parameters without correlations in a $\pm$ 30\% box
around the experimental values of $A_0$ and $A_2$ given by the cross hairs.
The figure shows the HV and NDR results, corresponding
to varying $\vev{\bar{q}q}$, $\vev{GG}$, $\Lambda^{(4)}_{\rm QCD}$ 
and $M$ in the ranges given in \eqs{range-hv}{range-ndr}.}
\label{diHV}
\end{figure}
%%%%%%%%%%%%%%%%%%%%%%%%%%%%%%%%%%%%%%%%%%%%%%%

\begin{figure}
\includegraphics[scale=0.5]{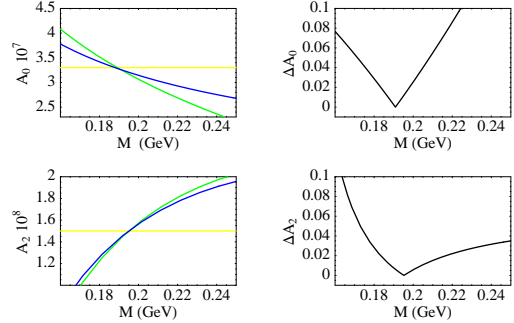}
\caption{Test of the $\gamma_5$-scheme stability of the $A_0$ and $A_2$
amplitudes as functions of $M$. 
The light (dark) curves correspond to the NDR (HV) results, 
while the horizontal lines mark the experimental values. 
The two figures on the right plot
$\Delta A_I \equiv 2 | (A_I^{HV}-A_I^{NDR})/(A_I^{HV}+A_I^{NDR})|$. 
For the values of \qq\ and \GG\ in \eqs{range-hv}{range-ndr} we find 
$\gamma_5$-scheme independence for $M\simeq 190-200$ MeV.}
\label{stab}
\end{figure}

Hadronic matrix elements in the $\chi$QM depend on the
$\gamma_5$-scheme utilized~\cite{trieste97}. Their dependence
partially cancels that of the short-distance NLO Wilson coefficients. 
Because this compensation is only numerical, and not analytical, we
take it as part of our phenomenological approach. A formal
$\gamma_5$-scheme matching can only come from a model more complete
than the $\chi$QM. Nevertheless, the result, as shown in
Fig.~\ref{stab}, is rather convincing.

By taking $$\Lambda^{(4)}_{\rm QCD} = 340 \pm 40\ \mbox{MeV}$$
and fitting at the scale $\mu = 0.8$ GeV the amplitudes in \eqs{A_0}{A_2}
to their experimental values, allowing for a $\pm$ 30\% 
systematic uncertainty, we find (see Fig.~\ref{diHV})
\bea
M & =&  195^{+25}_{-15} \; \mbox{MeV} \, ,
\nn \\
\langle \alpha_s GG/ \pi \rangle & = & 
\left( 330 \pm 5  \:\: \mbox{MeV} \right) ^4 \, ,
\nn \\
\langle \bar q q \rangle & = & \left(-235 \pm 25 \:\:\mbox{MeV} \right)^3\; .
\label{range-hv} 
\eea
in the HV-scheme,
and
\bea
M & = &  195^{+15}_{-10} \; \mbox{MeV} \, ,
\nn \\
\langle \alpha_s GG/ \pi \rangle & = &  
\left( 333^{+7}_{-6}  \:\: \mbox{MeV} \right) ^4 \, ,
\nn \\
\langle \bar q q \rangle & =& \left(-245 \pm 15 \:\:\mbox{MeV} \right)^3\; .
\label{range-ndr} 
\eea
in the NDR-scheme.

As shown by the light (NDR) and dark (HV) curves in Fig.~\ref{stab}, the
 $\gamma_5$-scheme dependence is controlled by the value of
$M$, the range of which is fixed thereby.
 The $\gamma_5$ scheme dependence of both amplitudes is minimized
for $M \simeq 190-200$ MeV.
The good $\gamma_5$-scheme stability is also enjoyed by \ratio\
and $\hat B_K$. 

%%%%%%%%%%%%%%%%%%%%%%% FIG.3 %%%%%%%%%%%%%%%%%
%%%%%%%%%%%%%%%%%%%%%%%%%%%%%%%%%%%%%%%%%%%%%%%

The fit of the amplitude $A_0$ and $A_2$ is obtained for
values of the quark and gluon
condensates which are in agreement with those
found in other approaches, i.e.\ QCD sum rules and lattice, 
although the relation between the gluon condensate of
QCD sum rules and lattice and that of the $\chi$QM is far from obvious. 
The value of the constituent quark mass $M$ is in good agreement
with that found by fitting radiative kaon decays~\cite{bijnens}.

\begin{figure}
\includegraphics[scale=0.4]{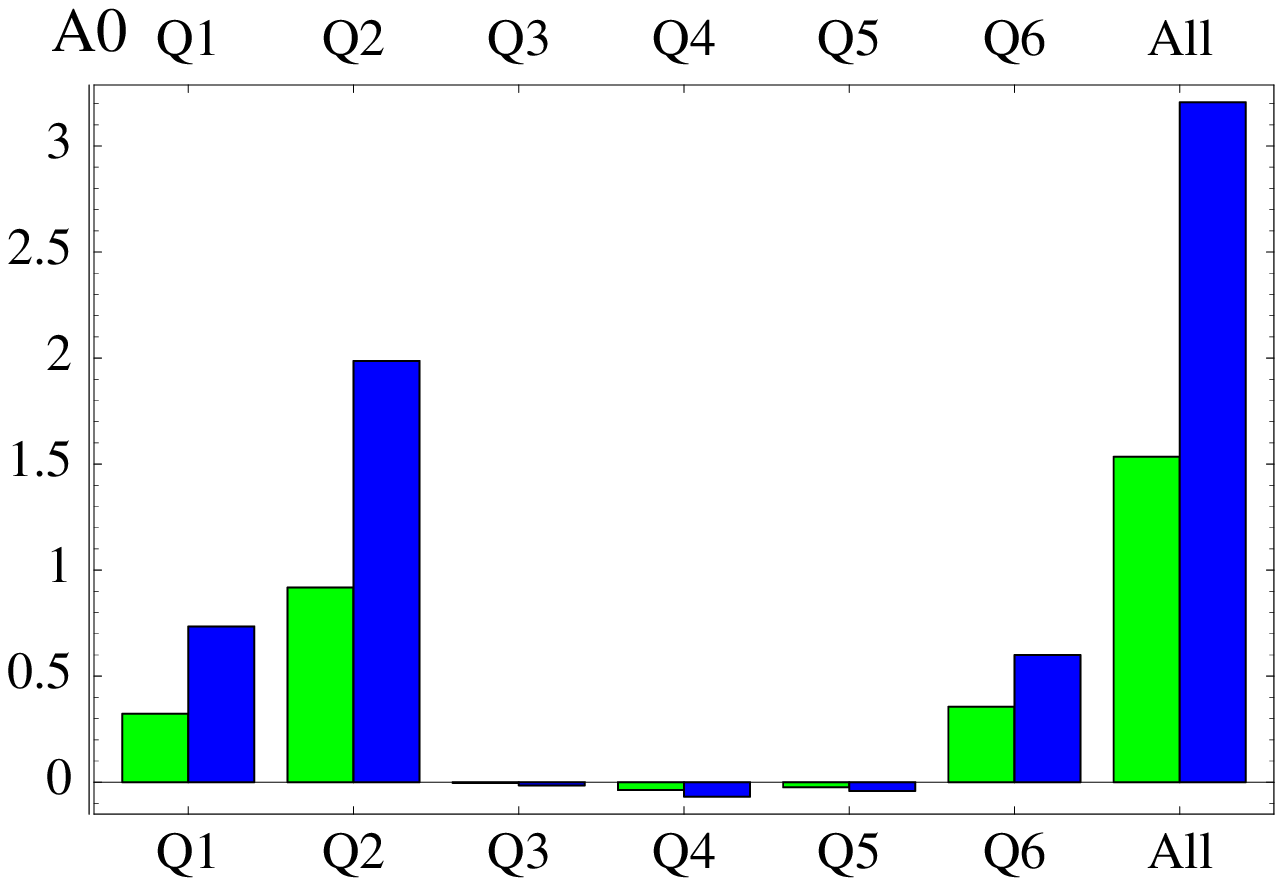}
\includegraphics[scale=0.4]{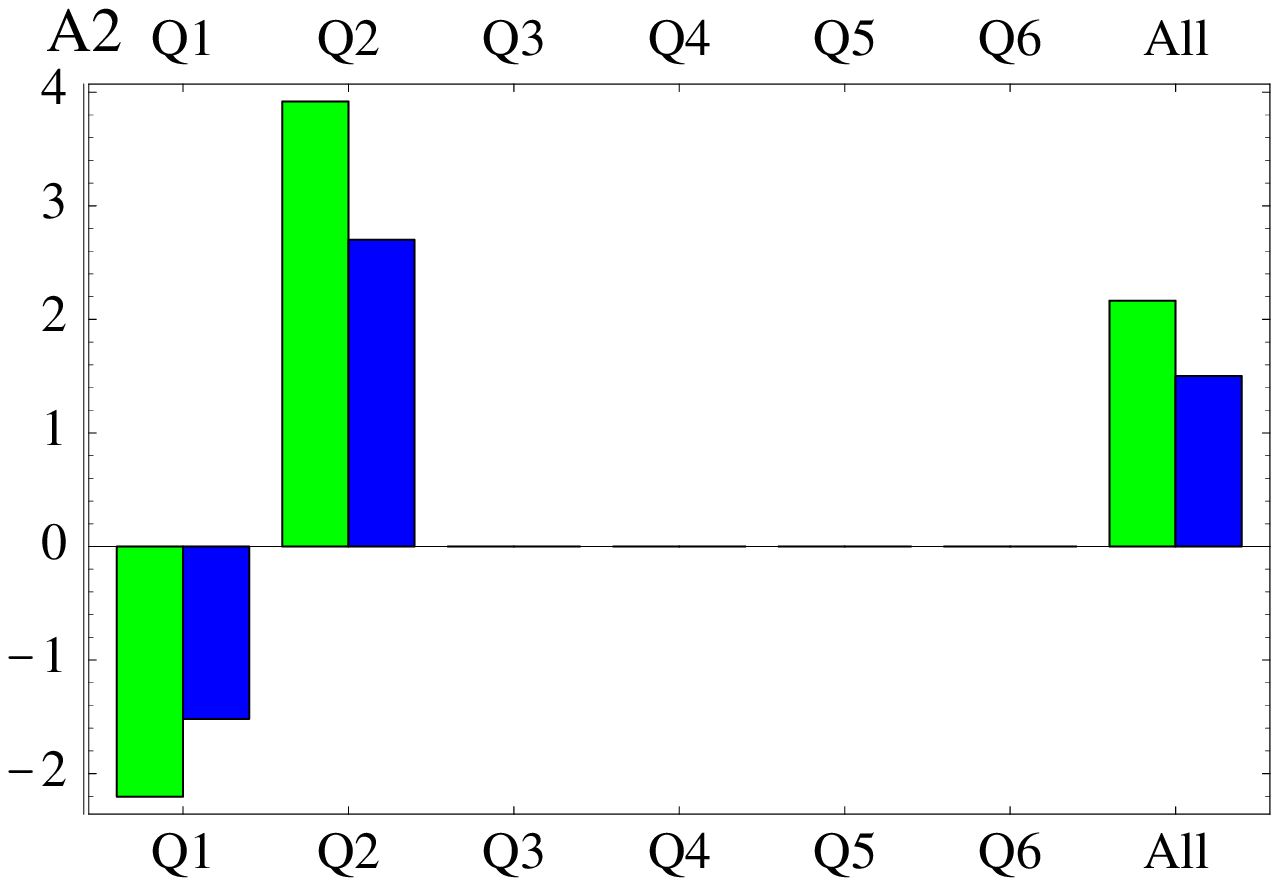}
\caption{Operator-by-operator contributions to $A_0 \times 10^7$
and $A_2 \times 10^8$ (GeV)
in the HV scheme. In  (light) dark the (LO) NLO value.
Notice the $O(p^4)$ enhancement of the gluon penguin operator
in the $I=0$ amplitude.}
\label{isto02}
\end{figure}

In Fig.~\ref{isto02} we present the anatomy of the relevant operator
contributions to the \CP\ conserving amplitudes. It is worth noticing
that, because of the NLO enhancement of the $I=0$ matrix elements
(mainly due to the chiral loops), the gluon penguin contribution
to $A_0$ amounts to about 20\% of the amplitude.

%%%%%%%%%%%%%%%%%%%%%%% FIG.4 %%%%%%%%%%%%%%%%%
%%%%%%%%%%%%%%%%%%%%%%%%%%%%%%%%%%%%%%%%%%%%%%%

Turning now to the $\Delta S =2$ Lagrangian,
by using the input values found by fitting the $\Delta I =1/2$ rule 
we obtain in both $\gamma_5$-schemes
\be
\hat B_K = 1.1 \pm 0.2 \, . \label{bk}
\ee
The result (\ref{bk}) includes chiral corrections up to $O(p^4)$ and it
agrees with what we found in~\cite{trieste97}. 

Notice that no estimate of \ratio\ can be considered complete unless
it also gives a value for $\hat B_K$. The case of the $\chi$QM, for
instance, is telling insofar as the enhancement of $B_6$ is partially
compensated for by a large $\hat B_K$ (and accordingly a smaller 
$\Im \lambda_t$).

\subsection{The bag parameters $B_i$}

The bag parameters $B_{i}$, defined as
\be
B_i \equiv \langle Q_i \rangle^{\rm model}/\langle Q_i \rangle^{\rm
VSA} \, ,
\ee 
have become a standard way of displaying the values of the hadronic
matrix elements in order to compare them among various
approaches. However they must be used with care because of their
dependence on the
renormalization scheme and scale, as well as on the choice of the VSA
parameters. 

\begin{small}
\begin{table}
\begin{tabular}{c|l|l}
 & HV & NDR \\
\hline
$B^{(0)}_1$  & $9.3$ & $9.7$ \\
$B^{(0)}_2$  & $2.8$ & $2.9$ \\
$B^{(2)}_1 =B^{(2)}_2 $  & $0.42$ &  $0.39$\\
$B_3$ & $-2.3$ & $-3.0$ \\
$B_4$ & 1.9 & 1.3\\
$B_5 \simeq B_6$ & $1.8 \times \frac{(-240\ {\rm MeV})^3}{\vev{\bar q q}}$ 
& $1.3 \times \frac{(-240\ {\rm MeV})^3}{\vev{\bar q q}}$  \\
$B_7^{(0)} \simeq B_8^{(0)}$ &$2.6$&$2.4$ \\
$B_9^{(0)}$ &$3.5$ & $3.4$ \\
$B_{10}^{(0)}$& $4.3$ & $5.2$ \\
$B_7^{(2)} \simeq B_8^{(2)}$ & $0.89$ & $0.84$ \\
$B_9^{(2)}=B_{10}^{(2)}$ & $0.42$ & $0.39$ \\
\end{tabular}
\caption{Central values of the $B_i$ factors in the HV and NDR
renormalization schemes. 
For $B_{5,6}$ the leading scaling dependence on \qq\ is explicitly
shown for a conventional value of the condensate. 
All other $B_i$ factors are either independent or very
weakly dependent on \qq. The dependence on \GG\ in the ranges 
of \eqs{range-hv}{range-ndr} remains always below 10\%.}
\label{bi}
\end{table}
\end{small}

They are given in the $\chi$QM in table \ref{bi} in the HV and NDR
schemes, at $\mu= 0.8$ GeV, for the central value of 
$\Lambda_{\rm QCD}^{(4)}$.
The uncertainty in the matrix elements of the penguin operators $Q_{5-8}$
arises from the variation of \qq. This affects mostly the $B_{5,6}$
parameters because of the leading
linear dependence on \qq\ of the $Q_{5,6}$ matrix elements in the $\chi$QM,
contrasted to the quadratic dependence of the corresponding VSA
matrix elements. Accordingly, $B_{5,6}$ scale as 
$\langle \bar q q \rangle ^{-1}$, or via PCAC as $m_q$, and 
therefore are sensitive to the value chosen for these parameters.
For this reason, we have reported the
corresponding values of $B_{5,6}$ when the quark condensate in the VSA
is fixed to its PCAC value.
It should however be stressed that such a dependence is not physical
and is introduced by the arbitrary normalization on the VSA result.
The estimate of \eprime\ is therefore almost independent of $m_q$,
which only enters the NLO corrections and the determination of $\hat B_K$.

The enhancement of the $Q_{5,6}$ matrix elements with respect to
the VSA values---the conventional normalization of the VSA matrix elements 
corresponds to taking 
\qq$(0.8\ {\rm GeV})\simeq (-220\ {\rm MeV})^3$---is mainly 
due to the NLO chiral loop contributions.
Such an enhancement, due to final state interactions, has been
found in $1/N_c$ analyses beyond LO~\cite{dortmund,BijnensPrades},
as well as old and recent dispersive studies~\cite{dispersive}.

\section{Bounds on $\Im \lambda_t$}

The updated measurements for the CKM elements $|V_{ub}/V_{cb}|$
 implies a change in the determination of the Wolfenstein
parameter $\eta$ that enter in $\Im \lambda_t$. This is of particular 
relevance because it affects proportionally the value of \ratio.

%%%%%%%%%%%%%%%%%%%%%%% FIG.5 %%%%%%%%%%%%%%%%%
\begin{figure}
\includegraphics[scale=0.4]{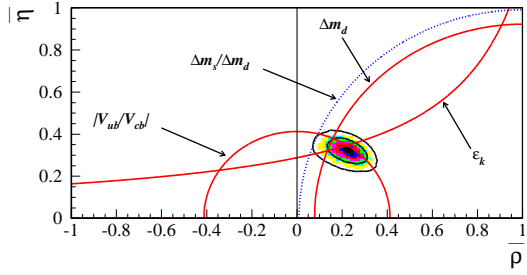}
\caption{Bounds on the Wolfenstein parameters $\bar \eta \equiv
(1-\lambda^2/2)$ 
and $\bar \rho \equiv (1-\lambda^2/2) $ and distribution of $\Im
\lambda_t$ according to Parodi et al. for the input parameters in 
\cite{trieste00}.}
\label{parodi}
\end{figure}
%%%%%%%%%%%%%%%%%%%%%%%%%%%%%%%%%%%%%%%%%%%%%%%

The allowed values for $\Im \lambda_t \simeq \eta |V_{us}||V_{cb}|^2$ 
are found by imposing the
experimental constraints for \eps,  $|V_{ub}/V_{cb}|$, $\Delta m_d$
and  $\Delta m_s$.
By using the method of Parodi, Roudeau and Stocchi~\cite{parodi}, who
have run their program starting from our  inputs listed in
\cite{trieste00}, it is found that
\be
\Im \lambda_t = (1.14 \pm 0.11) \times 10^{-4} \, , \label{imlamti}
\ee
where the error is determined by the 
Gaussian distribution in Fig.~\ref{parodi}. 

\section{Estimating \ratio\ }

\begin{figure}           
\includegraphics[scale=0.4]{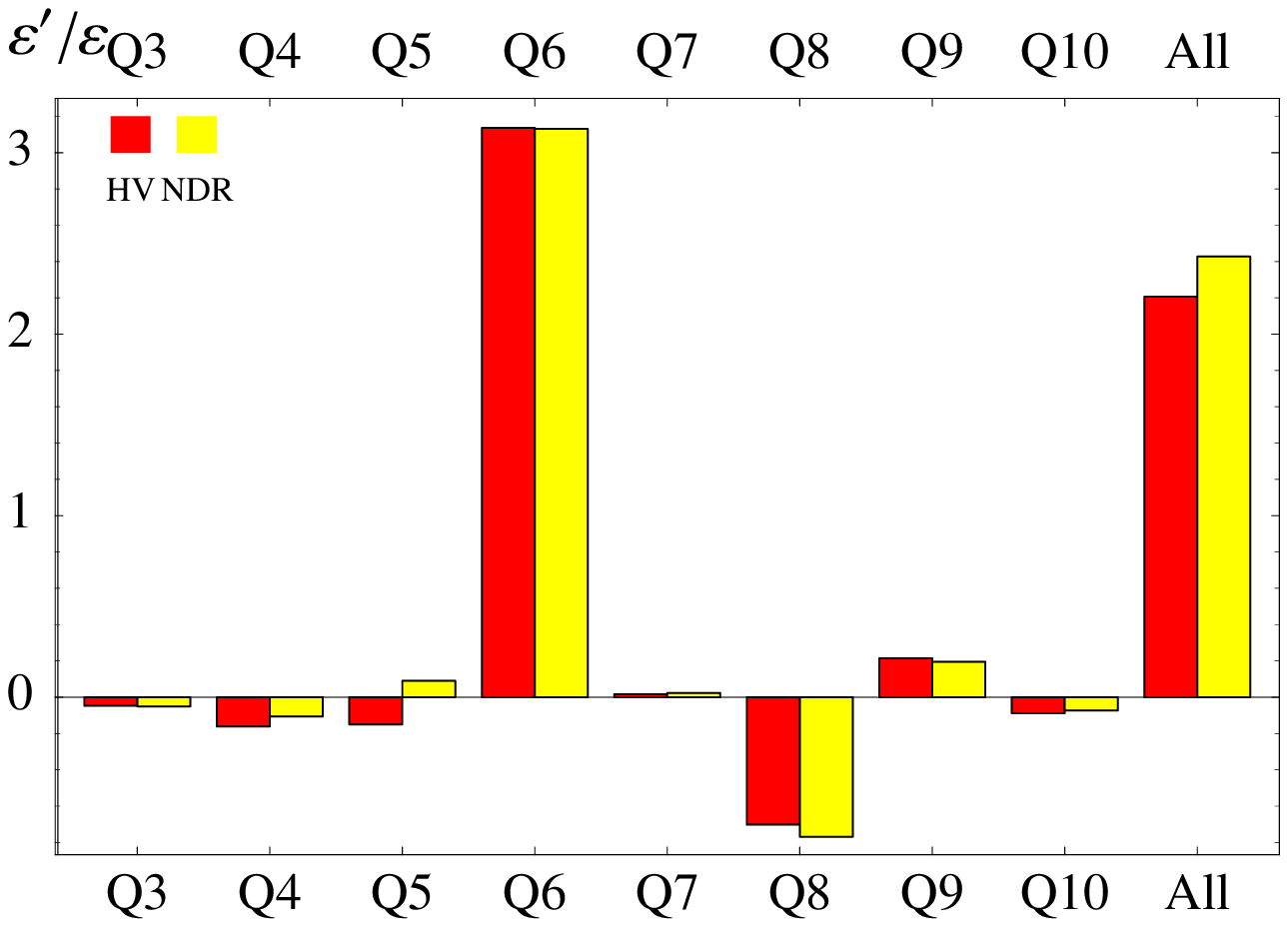}
\includegraphics[scale=0.4]{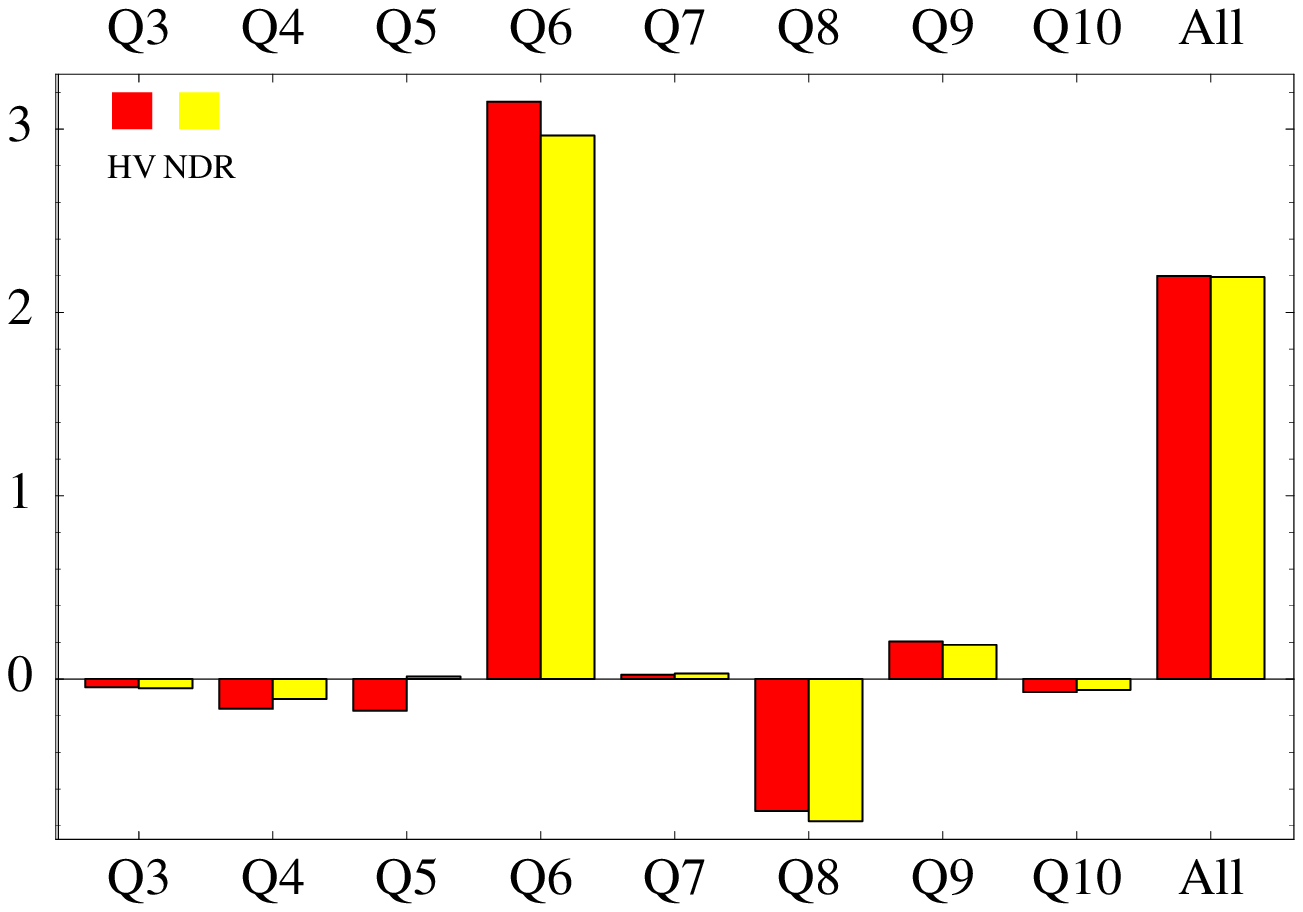}
\caption{Contribution to \ratio\  (in units of $10^{-3}$)
of each penguin operator in the HV and
NDR $\gamma_5$-schemes.
The figure on the left (right) corresponds to $\mu = 0.8$ ($1.0$) GeV.}
\label{histo}
\end{figure}

The value of \eprime\ computed by taking
all input parameters at their central values is shown in
Fig.~\ref{histo}. The figure shows
the contribution to \eprime\ of the various operators in two
$\gamma_5$ renormalization schemes at $\mu = 0.8$ GeV and $1.0$ GeV.
The advantage of such a histogram is that, contrary to the $B_i$, the
size of the individual contributions does not depend on some
conventional normalization.

As the histogram makes it clear, 
the gluon-penguin operator $Q_6$ dominates
and there is very little cancellation with the
electroweak penguin operator $Q_8$. 
The dominance of the $I=0$ components in the $\chi$QM 
originates from the $O(p^4)$ chiral corrections, 
the detailed size of which is determined
by the fit of the $\Delta I=1/2$ rule. 
It is a nice feature of the approach that the renormalization
scheme stability imposed on the \CP\ conserving amplitudes is numerically
preserved in \ratio. 
The comparison of the two figures shows also the remarkable
renormalization scale stability of the central value 
once the perturbative running of
the quark masses and the quark condensate is taken into account.

In what follows, the model-dependent parameters
$M$, \GG\ and \qq\ 
are uniformly varied in their given ranges (flat scanning),  
while the others are sampled according to their normal distributions.
Values of \ratio\ found in the HV and NDR schemes 
are included with equal weight.

%%%%%%%%%%%%%%%%%%%%%%% FIG.6 %%%%%%%%%%%%%%%%%
%%%%%%%%%%%%%%%%%%%%%%%%%%%%%%%%%%%%%%%%%%%%%%%

\begin{figure}           
\includegraphics[scale=0.5]{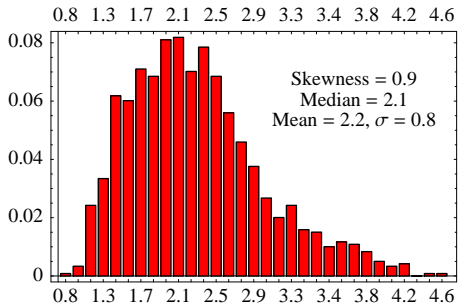}
\caption{Distribution of values of \ratio\ (in units of $10^{-3}$). 
Normalized bins are plotted against
the values of \ratio\ of each bin.}
\label{dist}
\end{figure}

For a given set, a distribution is
obtained by collecting the values of \ratio\ in bins of
a given range. This is shown in  Fig.~\ref{dist} for a particular
choice of bins.
Because the skewness of the distribution is less than one,
the mean and the standard deviation are a good estimate of the central
value and the dispersion of values around it.

%%%%%%%%%%%%%%%%%%%%%%% FIG.7 %%%%%%%%%%%%%%%%%
%%%%%%%%%%%%%%%%%%%%%%%%%%%%%%%%%%%%%%%%%%%%%%%

This statistical analysis yields 
\be
\varepsilon '/\varepsilon = (2.2 \pm 0.8) \times 10^{-3}
\, . \label{stretto}
\ee
A more conservative estimate of the uncertainties is obtained 
via the flat scanning of the input parameters, which gives 
\be
0.9\times 10^{-3} < \Re\: \varepsilon '/\varepsilon < 4.8\times 10^{-3}
\, . \label{largo}
\ee
In both estimates a theoretical systematic error of $\pm 30\%$ is included in
the fit of the \CP\ conserving amplitudes $A_0$ and $A_2$. 

The stability of the numerical outcomes is only marginally affected 
by shifts in the value of $\Omega_{\eta+\eta'}$.
Any effective variation of $\Omega_{\eta+\eta'}$ is
anti-correlated to the value of \GG\ obtained in the fit of $A_2$.
We have verified that this affects the calculation of $\hat B_K$ and
the consequent determination
of $\Im \lambda_t$ in such a way to compensate numerically 
in \ratio\ the change of $\Omega_{\eta+\eta'}$. 
Waiting for a confident assessment of the NLO isospin violating effects
in the $K\to\pi\pi$ amplitudes, we have used for
$\Omega_{\eta+\eta'}$ the `LO' value $0.25 \pm 0.10$.

The weak dependence on some poorly-known parameters is
a welcome outcome of the correlation among hadronic
matrix elements enforced in our semi-phenomenological approach
by the fit of the $\Delta I = 1/2$ rule.

While the $\chi QM$ approach to the hadronic matrix elements relevant
in the computation of \ratio\ has many advantages over other
techniques and has proved its value in the prediction 
of what has been then found in the experiments, it
has a severe short-coming insofar as the matching scale has to be
kept low, around 1 GeV and therefore the Wilson coefficients have to
be run down at scales that are at the limit of the applicability
of the renormalization-group equations. Moreover, the matching itself
suffers of ambiguities that have not be completely solved. For these
reasons we have insisted all along that the approach is
semi-phenomenological and that the above shortcomings are to be absorbed in
the values of the input parameters on which the fit to the \CP\
conserving amplitudes is based.
Because of its simplicity, the  $\chi$QM  is clearly not the final word
and it can now been abandoned---as a ladder used to climb a wall after we
are on the other side---as we work for better estimates, in
particular, those from the lattice simulations.

\section{Other estimates}

\begin{figure}            
\includegraphics[angle=0,scale=0.5]{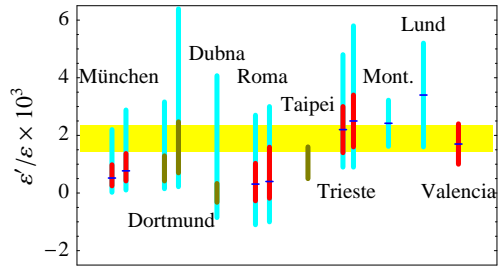}
\caption{Theory vs.\ experiment in the year 2000. 
The gray band is the
average experimental result. See \cite{munich99}, \cite{roma99},
\cite{dortmund}, \cite{dubna}, \cite{taipei}, \cite{monpellier}, 
\cite{lund}, \cite{valencia} 
for details on the various estimates.}
\label{expvsth}
\end{figure}

Figure~\ref{expvsth} summarizes the present status
of theory versus experiment. In addition to the current $\chi$QM calculation
(and an independent estimate similarly based on the $\chi$QM),
there are eight estimates of \ratio\ published in the last year.
Trieste's, M\"unchen's and Roma's ranges are updates of their
respective older estimates, while the others are altogether new.

%%%%%%%%%%%%%%%%%%%%%%% FIG.9 %%%%%%%%%%%%%%%%%
%%%%%%%%%%%%%%%%%%%%%%%%%%%%%%%%%%%%%%%%%%%%%%%

The estimates reported are discussed in \cite{trieste00}. The effects first
pointed out by the $\chi$QM have been taken up and refined by some of the
other approaches, in particular: non-factorizable 
corrections~\cite{dortmund,dubna,taipei,lund}
and FSI~\cite{dortmund,dubna,valencia}. 
Figure~\ref{expvsth} 
makes clear that, except for estimates heavily based on VSA for the
crucial operator $Q_6$, all other estimates
 are in agreement with the experiments.

\bigskip
\section*{Acknowledgments}

It is a pleasure to thank my collaborators S.~Bertolini and J.~O.~Eeg
and my former students V.~Antonelli and E.~I.~Lashin 
for the work done together.

%%%%%%%%%%%%%%%%%%%%%%%%%%%%%%%%%%%%%%%%%%%%%%%%%%%%%%%%%%%%%%%%%%%%%%%%
%------------------------- REFERENCES ------------------------------

%%%%%%%%%%%%%%%%%%%%%%%%%%%%%%%%%%%%%%%%%%%%%%%%%%%%%%%%%%%%%%%%%%%%%%%%%
\end{document}